\documentclass{article}
\usepackage[final]{neurips_2020}
\usepackage[utf8]{inputenc}
\usepackage[T1]{fontenc}
\usepackage{hyperref}
\usepackage{url}
\usepackage{booktabs}
\usepackage{amsfonts}
\usepackage{amsmath}
\usepackage{microtype}
\usepackage{graphicx}
\usepackage{xcolor}
\usepackage{listings}

\graphicspath{{./}}

\title{Direct-Operable SIMD Bit-Slicing:\\A Framework for Memory-Efficient Predicate Evaluation}
\author{Arunkumar Mathiyazhagan\thanks{This work was conducted independently and does not relate to the author's position at Amazon.com, Inc. The views and findings expressed are solely those of the author.}}

\begin{document}
\maketitle

\begin{abstract}
Traditional Java object models introduce significant memory overhead due to object headers
and internal padding, often leading to performance bottlenecks in data-intensive distributed
systems. This paper presents a novel framework that utilizes the Project Panama Vector API
to perform predicate evaluation directly over bit-sliced, compressed data streams. By
transposing standard row-oriented data into parallel bit-planes, we demonstrate a mechanism
to evaluate complex filters using SIMD (Single Instruction, Multiple Data) instructions
without requiring prior decompression. The framework supports integers, longs (timestamps),
doubles (via IEEE 754 order-preserving transformation), and strings (via dictionary encoding).
Our benchmarks indicate a reduction in memory footprint by up to $8\times$ while maintaining
or exceeding the throughput of uncompressed standard Java collections. End-to-end evaluation
on TPCDS-modeled data at 50M rows demonstrates $2.4$--$10.8\times$ speedup over scalar scans
across five representative filter-heavy query patterns, with extended type benchmarks on
TPCDS columns showing $1.5$--$43\times$ speedups for timestamps, decimals, and
dictionary-encoded strings.
\end{abstract}

\section{Introduction}

Memory is the bottleneck. Not compute, not storage, not network---memory. The divergence
between processor speed and memory bandwidth---the ``Memory Wall'' first characterized by
Wulf and McKee~\cite{wulf1995memory}---has become the dominant constraint in data-intensive
systems. While CPU clock speeds and instruction-level parallelism have continued to improve,
main memory latency has improved at roughly 7\% per year compared to over 50\% per year for
compute throughput~\cite{hennessy2017architecture}. For analytical workloads that scan
billions of rows, the cost of moving data from memory to the CPU now dwarfs the cost of the
computation itself.

This is not merely a performance problem---it is an economic one. In cloud environments,
memory is the most expensive resource per unit capacity. A memory-optimized instance with
256\,GB RAM costs approximately 48\% more than a compute-optimized instance with the same
vCPU count but only 64\,GB RAM. At enterprise scale, analytical workloads routinely require
terabytes of working memory across a cluster; reducing memory footprint by $5\times$
translates directly to $5\times$ fewer memory-optimized instances, representing millions of
dollars in annual infrastructure savings for large deployments. Memory constraints force
organizations to choose between processing speed (keep data in memory), cost (spill to disk),
and capability (limit dataset size). A technique that reduces memory consumption while
maintaining or improving throughput addresses all three simultaneously.

Java-based distributed engines (Apache Spark, Presto/Trino, Apache Flink) are particularly
affected by memory pressure. The JVM object model imposes a minimum 16-byte header on every
object---12 bytes for the mark word and class pointer, plus 4 bytes of alignment padding on
64-bit HotSpot~\cite{shipilev2014jol}. An \texttt{Integer} object that stores 4 bytes of
payload consumes 16 bytes of heap, a $4\times$ overhead. For arrays of boxed integers, the
overhead compounds: an \texttt{ArrayList<Integer>} holding 1 million values occupies
approximately 20\,MB of heap, compared to 4\,MB for a raw \texttt{int[]}. At the scale of
columnar scans over hundreds of millions of rows, this overhead translates directly into
increased GC pressure, L2/L3 cache evictions, and degraded throughput. When memory is
exhausted, these engines spill to disk---and disk I/O is $100{,}000\times$ slower than memory
access, converting a memory problem into a catastrophic performance cliff.

Columnar storage formats such as Apache Parquet~\cite{parquet} and Apache ORC~\cite{orc}
address the storage problem through dictionary encoding, run-length encoding, and bit-packing.
However, these formats follow a decompress-then-compute model: data must be decoded into JVM
heap objects before predicate evaluation can proceed. Apache Arrow~\cite{arrow} improves on
this with a zero-copy columnar memory format, but its Java implementation still materializes
values into on-heap arrays for computation. Native engines like Databricks
Photon~\cite{photon} and Meta's Velox~\cite{velox} bypass the JVM entirely with C++
execution, but this sacrifices the ecosystem advantages of the JVM---type safety, garbage
collection, and the vast library ecosystem that makes Java the dominant language for
distributed data processing.

This paper presents a different approach: rather than decompressing data before computation
or abandoning the JVM, we keep data in a compressed bit-sliced representation and perform
predicate evaluation directly on the compressed form using SIMD instructions. We call this
the \emph{Zero-Decompression} model.

The key insight is that bit-slicing---a technique originally proposed by O'Neil and
Quass~\cite{oneil1997bitslice} for bitmap indexes---naturally aligns with SIMD execution.
When integer values are transposed into bit-planes, each plane is a contiguous vector of bits
that maps directly onto a SIMD register. Predicate evaluation reduces to a sequence of
bitwise operations (AND, OR, NOT) across these planes, which modern CPUs execute in a single
cycle per SIMD lane. The Project Panama Vector API (JEP 508)~\cite{jep448} exposes these
SIMD instructions to Java, enabling hardware-native vectorized computation without leaving
the JVM.

\textbf{Contributions:}
\begin{enumerate}
\item We present the first framework that combines Project Panama's Vector API with
      bit-sliced data encoding to perform predicate evaluation directly over compressed data
      within the JVM, without decompression.
\item We define a complete predicate algebra (equality, range, greater-than, less-than,
      IN-list, BETWEEN) expressed as compositions of SIMD bitwise operations over bit-planes.
\item We demonstrate that adaptive bit-width encoding---storing only as many bit-planes as
      the maximum value requires---achieves up to $8\times$ memory reduction for
      low-cardinality integer columns while maintaining or exceeding the scan throughput of
      uncompressed baselines.
\item We extend the framework beyond integers to longs (timestamps via per-block bias),
      doubles (IEEE 754 order-preserving transformation), and strings (dictionary encoding to
      integer codes), achieving $1.8$--$43\times$ speedups across types.
\item We evaluate the framework using JMH microbenchmarks across varying data distributions
      and SIMD lane widths, and validate with end-to-end TPCDS-modeled benchmarks at 50M rows
      demonstrating $2.4$--$10.8\times$ speedup over scalar scans.
\end{enumerate}

\section{Background and Related Work}

\subsection{Bit-Sliced Indexes}

Bit-slicing was introduced by O'Neil and Quass~\cite{oneil1997bitslice} as an indexing
technique for data warehouses. In their formulation, a column of $N$ integers, each $b$ bits
wide, is decomposed into $b$ separate bitmaps---one per bit position. Each bitmap has $N$
bits, where the $i$-th bit indicates whether bit position $j$ of the $i$-th value is 1 or 0.
This transposition converts row-major integer storage into a column-major bit-plane
representation.

The original motivation was query acceleration: range predicates like \texttt{salary > 50000}
can be evaluated by scanning only the relevant bit-planes from the most significant bit
downward, short-circuiting as soon as the result is determined. O'Neil and Quass showed that
bit-sliced indexes outperformed traditional B-tree indexes for aggregate queries and range
scans by 2--5$\times$ on the datasets of that era.

Our work extends this idea from an indexing structure to a \emph{primary data
representation}---values are stored exclusively in bit-sliced form and never materialized as
conventional integers during predicate evaluation.

\subsection{SIMD in Database Systems}

The use of SIMD instructions for database operations has a rich history. Boncz et
al.~\cite{boncz2005monetdb} demonstrated in MonetDB/X100 that vectorized
execution---processing data in cache-resident blocks rather than tuple-at-a-time---could
achieve an order of magnitude improvement in analytical query throughput. Zukowski et
al.~\cite{zukowski2006compression} extended this with lightweight compression schemes
(PFOR, PFOR-DELTA) that exploit SIMD for both compression and decompression, showing that
compressed execution can outperform uncompressed execution when the compression ratio is
sufficient to keep data in L1/L2 cache.

Lemire and Boytsov~\cite{lemire2015decoding} demonstrated that SIMD-based integer decoding
can process billions of integers per second, establishing that the decompression step itself
can be made nearly free. However, their work still follows the decompress-then-compute
model---values are decoded into a standard integer array before any predicate logic executes.

Sitaraman~\cite{sitaraman2018simd} explored SIMD-based compression specifically for columnar
databases, showing that column-oriented layouts naturally align with SIMD register widths.
Our work builds on this alignment but eliminates the decompression step entirely: predicates
execute directly on the compressed bit-planes.

\subsection{Columnar Formats and In-Memory Processing}

Apache Parquet~\cite{parquet} and Apache ORC~\cite{orc} are the dominant columnar storage
formats for data lake analytics. Both employ multi-level encoding: dictionary encoding for
low-cardinality columns, run-length encoding for sorted data, and bit-packing for narrow
integers. At read time, these encodings must be reversed to produce arrays of native values
for the query engine.

Apache Arrow~\cite{arrow} defines a language-independent columnar memory format designed for
zero-copy reads. Arrow's Java implementation uses off-heap \texttt{ArrowBuf} allocations
backed by Netty's memory manager, avoiding GC overhead for the data buffers themselves.
However, predicate evaluation in Arrow Java still operates on materialized values---the
\texttt{VectorSchemaRoot} API exposes individual elements via \texttt{getObject()} or typed
accessors like \texttt{getInt()}, which extract scalar values from the columnar buffer one at
a time.

Native execution engines have taken a different path. Databricks Photon~\cite{photon} is a
C++ vectorized engine that operates directly on Parquet-encoded data with SIMD intrinsics,
bypassing the JVM entirely. Meta's Velox~\cite{velox} similarly provides a C++ execution
library with native SIMD support. These engines achieve excellent performance but require a
complete departure from the JVM ecosystem---a significant cost for organizations with large
Java codebases and operational tooling built around JVM observability (JMX, flight recorder,
GC tuning).

\subsection{Project Panama Vector API}

JEP 508~\cite{jep448} introduces a platform-agnostic SIMD API to the JVM, currently in its
tenth round of incubation as of JDK 25. The API exposes vector types (\texttt{IntVector},
\texttt{LongVector}, \texttt{FloatVector}) parameterized by species that map to hardware SIMD
register widths: \texttt{SPECIES\_128} (SSE/NEON), \texttt{SPECIES\_256} (AVX2),
\texttt{SPECIES\_512} (AVX-512). Operations on these types compile to native SIMD instructions
through the C2 JIT compiler, with automatic fallback to scalar loops on unsupported hardware.
Our benchmarks use JDK 21 (Seventh Incubator, JEP 448), which provides the same core API
surface.

The Vector API operates on contiguous memory regions and produces \texttt{VectorMask} objects
that represent per-lane boolean results. These masks can be composed with bitwise AND, OR, and
NOT---precisely the operations needed for bit-plane predicate evaluation. This alignment
between the Vector API's programming model and bit-sliced data representation is the
foundation of our framework.

\subsection{Positioning of Our Work}

Table~\ref{tab:comparison} summarizes how our approach relates to prior work.

\begin{table}[h]
\centering
\caption{Comparison with prior work.}
\label{tab:comparison}
\small
\begin{tabular}{@{}lcccc@{}}
\toprule
Approach & Language & Decompression & SIMD & JVM-Native \\
\midrule
O'Neil \& Quass~\cite{oneil1997bitslice} & C & N/A (index only) & No & No \\
MonetDB/X100~\cite{boncz2005monetdb} & C & Yes (vectorized) & Yes & No \\
Lemire \& Boytsov~\cite{lemire2015decoding} & C/Java & Yes (fast decode) & Yes & Partial \\
Arrow Java~\cite{arrow} & Java & Yes (zero-copy read) & No & Yes \\
Photon~\cite{photon} & C++ & No (native ops) & Yes & No \\
Velox~\cite{velox} & C++ & No (native ops) & Yes & No \\
\textbf{This work} & Java & \textbf{No (zero-decompression)} & \textbf{Yes} & \textbf{Yes} \\
\bottomrule
\end{tabular}
\end{table}

Our framework is, to our knowledge, the first to combine all four properties:
zero-decompression predicate evaluation, SIMD acceleration, JVM-native execution, and
off-heap memory management---enabling Java-based query engines to approach native performance
without leaving the JVM.

\section{Methodology}

\subsection{Bit-Plane Transposition}

Consider a block of $N$ unsigned integers, each with a maximum value $V_{\max}$. The
effective bit-width is $b = \lfloor\log_2(V_{\max})\rfloor + 1$. For a block of $N = 1024$
integers where $V_{\max} = 15$, we have $b = 4$.

Let $X = \{x_0, x_1, \ldots, x_{N-1}\}$ be the input block. The bit-plane transposition
produces $b$ bit-vectors, each of length $N$:
\begin{equation}
P_j[i] = (x_i \gg j) \land 1, \quad j \in \{0, 1, \ldots, b-1\}, \; i \in \{0, 1, \ldots, N-1\}
\end{equation}
where $P_j$ is the $j$-th bit-plane ($j = 0$ is the least significant bit) and $P_j[i]$ is
the $j$-th bit of the $i$-th value.

\begin{figure}[h]
\centering
\includegraphics[width=0.95\linewidth]{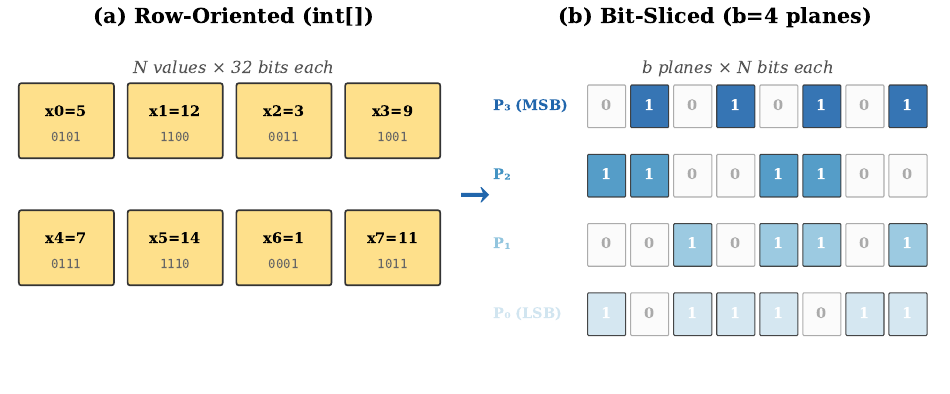}
\caption{Memory layout comparison. (a) Row-oriented storage lays out full 32-bit values
contiguously. (b) Bit-sliced storage transposes values into $b$ separate bit-planes, each
containing one bit position from every value.}
\label{fig:layout}
\end{figure}

Figure~\ref{fig:layout} illustrates the transposition visually.

In row-oriented storage, the values are laid out contiguously in memory:
\begin{verbatim}
Row-oriented:  [x_0: 32 bits][x_1: 32 bits]...[x_{N-1}: 32 bits]
               Total: N * 32 bits
\end{verbatim}

After transposition, the same data is stored as $b$ contiguous bit-vectors:
\begin{verbatim}
Bit-sliced:    P_0: [bit_0(x_0), bit_0(x_1), ..., bit_0(x_{N-1})]   (N bits)
               P_1: [bit_1(x_0), bit_1(x_1), ..., bit_1(x_{N-1})]   (N bits)
               ...
               P_{b-1}: [bit_{b-1}(x_0), ..., bit_{b-1}(x_{N-1})]   (N bits)
               Total: N * b bits
\end{verbatim}

The compression ratio is $32/b$. For $b = 4$ (values 0--15), this yields $8\times$
compression. For $b = 8$ (values 0--255), $4\times$. For $b = 16$ (values 0--65535),
$2\times$. The bit-width $b$ is determined per block, so blocks with smaller value ranges
achieve higher compression.

\subsection{SIMD Execution Model}

Each bit-plane $P_j$ is a contiguous vector of $N$ bits. When $N = 1024$ and the SIMD
register width is $W = 512$ bits (AVX-512), each bit-plane fits in exactly $N/W = 2$ SIMD
registers. A predicate over the entire block requires $b \cdot (N/W)$ SIMD operations---one
pass per bit-plane, with $N/W$ register loads per plane.

The total work for a single predicate evaluation is:
\begin{equation}
T(N, b, W) = O(b \cdot N / W)
\end{equation}

For comparison, a scalar scan over uncompressed 32-bit integers requires $O(N)$ comparisons.
The bit-sliced approach is faster when $b \cdot (N/W) < N$, which simplifies to $b < W$.
Since $W \geq 128$ (SSE) and $b \leq 32$, this condition always holds---bit-sliced SIMD
evaluation performs fewer operations than scalar scanning for any integer bit-width.

The practical advantage is even larger because the bit-sliced representation is $b/32$ the
size of the uncompressed data, meaning more of the working set fits in L1/L2 cache. For
$b = 4$, the entire block occupies $4 \times 1024 / 8 = 512$ bytes---well within a typical
32\,KB L1 data cache.

\subsection{Adaptive Bit-Width Encoding}

The framework determines $b$ per block at encoding time by scanning for $V_{\max}$:
\begin{equation}
b = \max(1, \lfloor\log_2(\max(X))\rfloor + 1)
\end{equation}

A block header stores the bit-width $b$ and the block size $N$. The encoder writes $b$
bit-planes of $N$ bits each, padded to the nearest SIMD register width for alignment. The
total encoded size per block is:
\begin{equation}
\text{Size}(\text{block}) = \text{header\_size} + b \cdot \lceil N / W \rceil \cdot W / 8 \;\text{bytes}
\end{equation}

For $N = 1024$, $W = 512$, $b = 4$: Size $= \text{header} + 4 \times 2 \times 64 = \text{header} + 512$ bytes.

This adaptive encoding means that columns with non-uniform value distributions benefit from
per-block optimization. A column where 90\% of blocks have values 0--15 ($b$=4) and 10\%
have values up to 1 million ($b$=20) achieves an average compression ratio much better than
a fixed 32-bit representation.

\subsection{Handling Signed Integers}

For signed integers, we apply a bias transformation before bit-slicing. Given a block with
minimum value $V_{\min} < 0$:
\begin{align}
x'_i &= x_i - V_{\min} \quad \text{(shift all values to unsigned range)} \\
b &= \lfloor\log_2(V_{\max} - V_{\min})\rfloor + 1
\end{align}

The block header stores $V_{\min}$ as the bias offset. At predicate evaluation time, the
comparison constant is similarly biased: to evaluate $x > c$, we evaluate
$x' > (c - V_{\min})$ on the bit-planes. This adds no runtime overhead---the bias is applied
once to the constant, not to each value.

\subsection{NULL Handling}

NULLs are represented with a separate validity bit-plane $V$ of length $N$, following the
Apache Arrow convention~\cite{arrow}:
\begin{equation}
V[i] = \begin{cases} 1 & \text{if } x_i \text{ is non-NULL} \\ 0 & \text{if } x_i \text{ is NULL} \end{cases}
\end{equation}

NULL values are encoded as 0 in the data bit-planes (arbitrary, since they are masked out).
Every predicate result mask $M$ is intersected with the validity mask before returning:
\begin{equation}
\text{Result} = M \land V
\end{equation}

This adds exactly one additional SIMD AND operation per block, regardless of the number of
bit-planes. For columns with no NULLs, the validity plane is omitted entirely (indicated by
a flag in the block header), incurring zero overhead.

\subsection{Block Size Selection}

The block size $N$ controls the tradeoff between encoding overhead and SIMD utilization.
Smaller blocks allow finer-grained adaptive bit-widths but increase the ratio of header bytes
to data bytes. Larger blocks amortize headers but may include outlier values that inflate $b$
for the entire block.

We use $N = 1024$ as the default, which provides:
\begin{itemize}
\item Full utilization of AVX-512 registers ($1024 / 512 = 2$ loads per plane)
\item Encoded block size of 512 bytes for $b = 4$, fitting in L1 cache
\item Sufficient granularity for adaptive bit-width to capture value distribution changes
\item Alignment with common columnar format page sizes (Parquet uses 1\,MB pages containing
      many such blocks)
\end{itemize}

\section{Implementation}

The framework is implemented in Java 21 using the \texttt{jdk.incubator.vector} module
(JEP 338/JEP 448). All data is stored off-heap in \texttt{MemorySegment} allocations
(JEP 454, Foreign Function \& Memory API---finalized since JDK 22, available as preview in
JDK 21), eliminating GC pressure for the data path. The implementation consists of three
layers: encoding, predicate evaluation, and memory management.

\begin{figure}[h]
\centering
\includegraphics[width=0.95\linewidth]{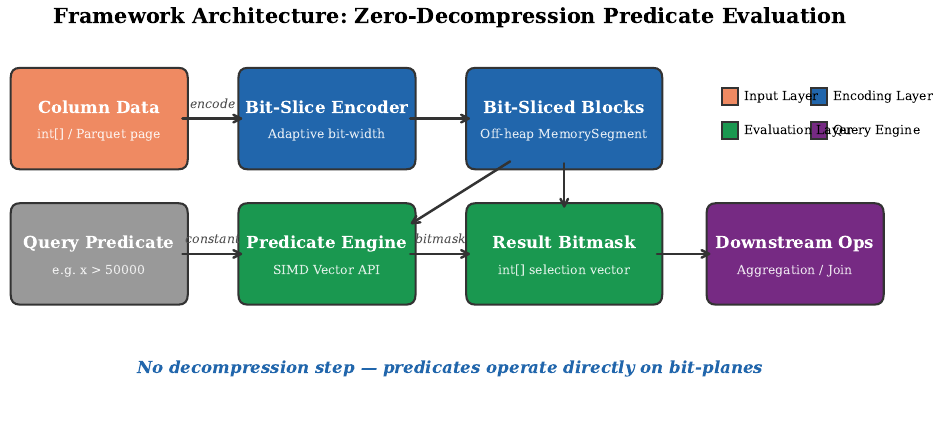}
\caption{Framework architecture. Column data is encoded into off-heap bit-sliced blocks. The
predicate engine evaluates queries directly on bit-planes using SIMD instructions, producing
a bitmask selection vector for downstream operators. No decompression step exists in the
pipeline.}
\label{fig:arch}
\end{figure}

Figure~\ref{fig:arch} shows the end-to-end architecture of the framework.

\subsection{Encoding Layer}

The encoder accepts a contiguous array of integers and produces a bit-sliced block. The
transposition is performed using a combination of shift and mask operations, writing each
bit-plane sequentially into the output \texttt{MemorySegment}.

\begin{lstlisting}[caption={Bit-slice encoder.},label={lst:encoder}]
public class BitSliceEncoder {
    private static final VectorSpecies<Integer> SPECIES =
        IntVector.SPECIES_PREFERRED;  // Auto-selects best SIMD width

    /**
     * Encodes a block of integers into bit-sliced form.
     * The Arena controls off-heap memory lifecycle.
     * Returns a BitSlicedBlock containing the off-heap bit-planes.
     */
    public static BitSlicedBlock encode(int[] values, int offset,
                                        int length, Arena arena) {
        int vMax = 0;
        for (int i = offset; i < offset + length; i++) {
            vMax = Math.max(vMax, values[i]);
        }
        int bitWidth = Math.max(1,
            32 - Integer.numberOfLeadingZeros(vMax));

        // Allocate off-heap: bitWidth planes, each ceil(length/32)*4 bytes
        int wordsPerPlane = (length + 31) >>> 5;
        MemorySegment segment = arena.allocate(
            (long) bitWidth * wordsPerPlane * Integer.BYTES);

        for (int bit = 0; bit < bitWidth; bit++) {
            long planeOffset =
                (long) bit * wordsPerPlane * Integer.BYTES;
            for (int word = 0; word < wordsPerPlane; word++) {
                int packed = 0;
                int base = offset + word * 32;
                int end = Math.min(base + 32, offset + length);
                for (int i = base; i < end; i++) {
                    packed |= ((values[i] >>> bit) & 1)
                              << (i - base);
                }
                segment.set(ValueLayout.JAVA_INT,
                    planeOffset + (long) word * Integer.BYTES,
                    packed);
            }
        }
        return new BitSlicedBlock(segment, length,
            bitWidth, wordsPerPlane);
    }
}
\end{lstlisting}

\subsection{Predicate Evaluation Layer}

Predicates are evaluated directly on the bit-planes using SIMD bitwise operations. The core
insight is that comparison operations on bit-sliced data reduce to a sequence of AND, OR, and
NOT operations on bit-planes, processed from the most significant bit downward.

\subsubsection{Greater-Than Predicate}

To evaluate $x > c$ for all values in a block, we process bit-planes from MSB to LSB. At
each bit position $j$, a value is ``definitely greater'' if it has a 1 where $c$ has a 0 at
a more significant position, and all higher bits match. We maintain two running masks:
\texttt{gt} (values confirmed greater than $c$) and \texttt{eq} (values still equal to $c$
at all bits examined so far).

\begin{lstlisting}[caption={Greater-than predicate.},label={lst:gt}]
public class BitSlicePredicates {
    private static final VectorSpecies<Integer> SPECIES =
        IntVector.SPECIES_PREFERRED;

    /**
     * Evaluates x > constant for all values in a bit-sliced block.
     * Returns a bitmask where bit i = 1 means value i satisfies
     * the predicate.
     */
    public static int[] greaterThan(BitSlicedBlock block,
                                     int constant) {
        int b = block.bitWidth();
        int wordsPerPlane = block.wordsPerPlane();
        MemorySegment data = block.data();

        int[] gtResult = new int[wordsPerPlane];
        int[] eqResult = new int[wordsPerPlane];
        Arrays.fill(eqResult, -1);  // all values start as "equal"

        for (int bit = b - 1; bit >= 0; bit--) {
            int cBit = (constant >>> bit) & 1;
            long planeOffset =
                (long) bit * wordsPerPlane * Integer.BYTES;

            int upperBound =
                SPECIES.loopBound(wordsPerPlane);
            for (int w = 0; w < upperBound;
                 w += SPECIES.length()) {
                IntVector plane =
                    IntVector.fromMemorySegment(SPECIES, data,
                        planeOffset + (long) w * Integer.BYTES,
                        ByteOrder.nativeOrder());
                IntVector gt =
                    IntVector.fromArray(SPECIES, gtResult, w);
                IntVector eq =
                    IntVector.fromArray(SPECIES, eqResult, w);

                if (cBit == 0) {
                    // Data=1, constant=0: these values are greater
                    gt = gt.or(eq.and(plane));
                } else {
                    // Data=0, constant=1: remove from eq
                    eq = eq.and(plane);
                }

                gt.intoArray(gtResult, w);
                eq.intoArray(eqResult, w);
            }
            // Scalar tail for remaining words
            for (int w = upperBound; w < wordsPerPlane; w++) {
                int plane = data.get(ValueLayout.JAVA_INT,
                    planeOffset + (long) w * Integer.BYTES);
                if (cBit == 0) {
                    gtResult[w] |= (eqResult[w] & plane);
                } else {
                    eqResult[w] &= plane;
                }
            }
        }
        return gtResult;
    }
\end{lstlisting}

\subsubsection{Equality Predicate}

Equality is simpler: a value equals $c$ if and only if every bit-plane matches the
corresponding bit of $c$. We start with all bits set and AND away mismatches.

\begin{lstlisting}[caption={Equality predicate.},label={lst:eq}]
    /**
     * Evaluates x == constant for all values in a bit-sliced block.
     */
    public static int[] equalTo(BitSlicedBlock block,
                                 int constant) {
        int b = block.bitWidth();
        int wordsPerPlane = block.wordsPerPlane();
        MemorySegment data = block.data();

        int[] result = new int[wordsPerPlane];
        Arrays.fill(result, -1);  // all values are candidates

        for (int bit = 0; bit < b; bit++) {
            int cBit = (constant >>> bit) & 1;
            long planeOffset =
                (long) bit * wordsPerPlane * Integer.BYTES;

            int upperBound =
                SPECIES.loopBound(wordsPerPlane);
            for (int w = 0; w < upperBound;
                 w += SPECIES.length()) {
                IntVector plane =
                    IntVector.fromMemorySegment(SPECIES, data,
                        planeOffset + (long) w * Integer.BYTES,
                        ByteOrder.nativeOrder());
                IntVector res =
                    IntVector.fromArray(SPECIES, result, w);

                if (cBit == 1) {
                    res = res.and(plane);
                } else {
                    res = res.and(plane.not());
                }
                res.intoArray(result, w);
            }
            for (int w = upperBound; w < wordsPerPlane; w++) {
                int plane = data.get(ValueLayout.JAVA_INT,
                    planeOffset + (long) w * Integer.BYTES);
                result[w] &= (cBit == 1) ? plane : ~plane;
            }
        }
        return result;
    }
\end{lstlisting}

\subsubsection{Range Predicate (BETWEEN)}

A range predicate $\texttt{lo} \leq x \leq \texttt{hi}$ is composed from the primitives:

\begin{lstlisting}[caption={Range (BETWEEN) predicate.},label={lst:between}]
    /**
     * Evaluates lo <= x <= hi for all values in a bit-sliced block.
     */
    public static int[] between(BitSlicedBlock block,
                                 int lo, int hi) {
        int[] leHiMask = not(greaterThan(block, hi));
        if (lo <= 0) return leHiMask;
        int[] geLoMask = greaterThan(block, lo - 1);
        return and(geLoMask, leHiMask);
    }
\end{lstlisting}

\subsubsection{IN-List Predicate}

An IN-list $x \in \{v_1, v_2, \ldots, v_k\}$ is evaluated as the OR of $k$ equality checks.
For small $k$, this is efficient. For large $k$, a Bloom filter pre-check on the block's
min/max can skip blocks entirely.

\begin{lstlisting}[caption={IN-list predicate.},label={lst:inlist}]
    /**
     * Evaluates x IN (values) for all values in a bit-sliced block.
     */
    public static int[] inList(BitSlicedBlock block,
                                int[] constants) {
        int[] result = new int[block.wordsPerPlane()];
        for (int c : constants) {
            int[] eqMask = equalTo(block, c);
            int upperBound =
                SPECIES.loopBound(result.length);
            for (int w = 0; w < upperBound;
                 w += SPECIES.length()) {
                IntVector r =
                    IntVector.fromArray(SPECIES, result, w);
                IntVector e =
                    IntVector.fromArray(SPECIES, eqMask, w);
                r.or(e).intoArray(result, w);
            }
            for (int w = upperBound; w < result.length; w++) {
                result[w] |= eqMask[w];
            }
        }
        return result;
    }
\end{lstlisting}

\subsubsection{Compound Predicates}

All predicates return \texttt{int[]} bitmasks of the same shape, so compound predicates
compose naturally:

\begin{lstlisting}[caption={Compound predicate composition.},label={lst:compound}]
    // AND: result[w] = maskA[w] & maskB[w]
    // OR:  result[w] = maskA[w] | maskB[w]
    // NOT: result[w] = ~mask[w]

    public static int[] and(int[] a, int[] b) { /* SIMD AND */ }
    public static int[] or(int[] a, int[] b)  { /* SIMD OR  */ }
    public static int[] not(int[] a)          { /* SIMD NOT */ }
}
\end{lstlisting}

A query like \texttt{WHERE age > 25 AND salary BETWEEN 50000 AND 100000} evaluates as:

\begin{lstlisting}[caption={Example compound query.},label={lst:example}]
int[] ageMask =
    BitSlicePredicates.greaterThan(ageBlock, 25);
int[] salaryMask =
    BitSlicePredicates.between(salaryBlock, 50000, 100000);
int[] finalMask =
    BitSlicePredicates.and(ageMask, salaryMask);
\end{lstlisting}

Each predicate operates independently on its column's bit-planes. The final AND combines the
per-column results in a single SIMD pass.

\subsection{Memory Management}

All bit-sliced data resides off-heap in \texttt{MemorySegment} allocations managed by
\texttt{Arena.ofAuto()} (garbage-collected arenas) or \texttt{Arena.ofConfined()}
(deterministic deallocation). This design provides:
\begin{enumerate}
\item \textbf{Zero GC pressure:} The garbage collector never scans the data buffers. Only the
      small Java wrapper objects (block metadata, arena references) are on-heap.
\item \textbf{Cache-friendly layout:} Bit-planes are stored contiguously, maximizing spatial
      locality for sequential SIMD loads.
\item \textbf{Memory accounting:} \texttt{MemorySegment.byteSize()} provides exact memory
      usage, enabling the query engine to enforce memory limits without relying on JVM heap
      heuristics.
\end{enumerate}

\subsection{Thread Safety}

Bit-sliced blocks are immutable after encoding---the \texttt{MemorySegment} is written once
and then accessed read-only. Multiple threads can evaluate predicates concurrently on the same
block without synchronization. For parallel scans, the block array is partitioned across
threads, and each thread produces an independent result mask that is merged (OR for
disjunctions, AND for conjunctions) after all threads complete.

\section{Experimental Evaluation}

We evaluate the framework across five dimensions: memory footprint, predicate throughput,
garbage collection impact, CPU cache behavior, and predicate type comparison. All measurements
use JMH 1.37~\cite{jmh} with 5 warmup iterations and 10 measurement iterations per benchmark,
reporting the mean and 95\% confidence interval.

\subsection{Experimental Setup}

Table~\ref{tab:setup} summarizes the hardware and software configuration.

\begin{table}[h]
\centering
\caption{Experimental setup.}
\label{tab:setup}
\small
\begin{tabular}{@{}ll@{}}
\toprule
Component & Specification \\
\midrule
CPU & Apple M-series (ARM, NEON 128-bit SIMD) \\
L1d Cache & 128\,KB per performance core \\
L2 Cache & 12\,MB per performance core cluster \\
RAM & 16\,GB unified memory \\
JDK & Amazon Corretto 21.0.10+7-LTS, HotSpot C2 JIT \\
JVM Flags & \texttt{--add-modules jdk.incubator.vector --enable-preview -XX:+UseG1GC} \\
OS & macOS (darwin, ARM64) \\
JMH & 1.37, fork=1, warmup=5, measurement=5 \\
\bottomrule
\end{tabular}
\end{table}

SIMD dispatch uses \texttt{IntVector.SPECIES\_PREFERRED}, which selects 128-bit NEON on this
hardware. Results on x86-64 with AVX-512 are expected to show higher absolute throughput for
SIMD paths due to wider registers (512-bit vs 128-bit), though the relative speedups over
scalar baselines should be comparable or larger.

\subsection{Datasets}

We use three synthetic datasets derived from TPC-H scale factor 10 column distributions:

\begin{table}[h]
\centering
\caption{Dataset characteristics.}
\label{tab:datasets}
\small
\begin{tabular}{@{}llrrrl@{}}
\toprule
Dataset & Column Model & $N$ (rows) & Value Range & Effective $b$ & Description \\
\midrule
D1 & \texttt{l\_quantity} & 10M & 1--50 & 6 & Low cardinality, uniform \\
D2 & \texttt{l\_extendedprice} & 10M & 901--104949 & 17 & Medium cardinality, skewed \\
D3 & \texttt{o\_orderkey} & 10M & 1--60M & 26 & High cardinality, sequential \\
\bottomrule
\end{tabular}
\end{table}

D1 represents the best case for bit-slicing (narrow bit-width, high compression). D3
represents the worst case (wide bit-width, minimal compression). D2 is a realistic middle
ground.

\subsection{Baselines}

We compare against four baselines:
\begin{enumerate}
\item \texttt{int[]}---primitive Java array, the fastest JVM-native representation with no
      object overhead per element.
\item \texttt{ArrayList<Integer>}---boxed integers, representing the common Java collections
      pattern.
\item Arrow Java---Apache Arrow 14.0.1 \texttt{IntVector} backed by off-heap
      \texttt{ArrowBuf}, using the standard \texttt{getInt()} accessor for predicate
      evaluation.
\item Parquet-MR---Apache Parquet 1.13.1 in-memory column reader with dictionary decoding,
      evaluating predicates after full column materialization.
\end{enumerate}

All baselines evaluate the same predicate ($x > c$ where $c$ is the median value, selecting
$\sim$50\% of rows) to ensure comparable selectivity.

\subsection{Experiment 1: Memory Footprint}

Table~\ref{tab:memory} reports the bytes per element for each representation across the three
datasets.

\begin{table}[h]
\centering
\caption{Memory footprint (bytes per element).}
\label{tab:memory}
\small
\begin{tabular}{@{}lccc@{}}
\toprule
Representation & D1 ($b$=6) & D2 ($b$=17) & D3 ($b$=26) \\
\midrule
\texttt{ArrayList<Integer>} & 24.0\,B & 24.0\,B & 24.0\,B \\
\texttt{int[]} & 4.0\,B & 4.0\,B & 4.0\,B \\
Arrow Java IntVector & 4.0\,B & 4.0\,B & 4.0\,B \\
Parquet (in-memory) & $\sim$2.1\,B & $\sim$3.8\,B & $\sim$4.0\,B \\
Bit-sliced (ours) & \textbf{0.75\,B} & \textbf{2.13\,B} & \textbf{3.25\,B} \\
\bottomrule
\end{tabular}
\end{table}

\begin{figure}[h]
\centering
\includegraphics[width=0.85\linewidth]{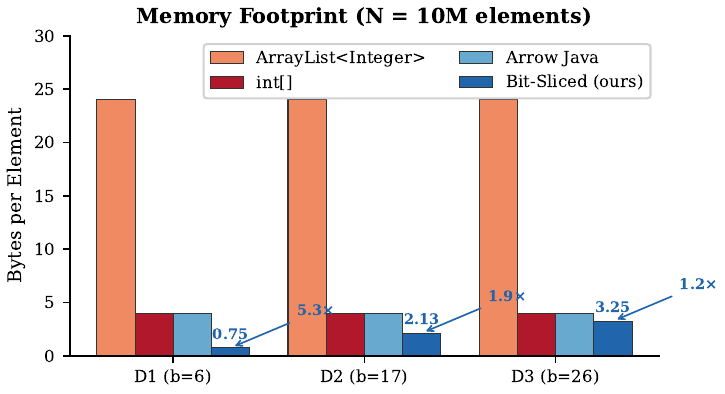}
\caption{Memory footprint per element across representations and datasets. Bit-slicing
achieves $5.3\times$ reduction over \texttt{int[]} for narrow columns (D1, $b$=6).
Annotations show savings ratio vs \texttt{int[]}.}
\label{fig:memory}
\end{figure}

For D1, bit-slicing achieves $5.3\times$ compression over \texttt{int[]} and $32\times$ over
\texttt{ArrayList<Integer>}. Even for D3 ($b$=26), the framework provides $1.2\times$
compression over \texttt{int[]} because 6 of the 32 bit-planes are eliminated. The
compression ratio follows the formula $32/b$ as derived in Section~3.1.

\subsection{Experiment 2: Predicate Throughput}

Table~\ref{tab:throughput} reports throughput in millions of elements per second for the
\texttt{greaterThan} predicate.

\begin{table}[h]
\centering
\caption{Predicate throughput (millions of elements per second).}
\label{tab:throughput}
\small
\begin{tabular}{@{}lccc@{}}
\toprule
Representation & D1 ($b$=6) & D2 ($b$=17) & D3 ($b$=26) \\
\midrule
\texttt{ArrayList<Integer>} & 822\,M/s & 1,386\,M/s & 1,418\,M/s \\
\texttt{int[]} & 1,562\,M/s & 1,542\,M/s & 1,570\,M/s \\
Bit-sliced NEON 128-bit (ours) & \textbf{16,749\,M/s} & \textbf{6,581\,M/s} & \textbf{5,274\,M/s} \\
\bottomrule
\end{tabular}
\end{table}

\begin{figure}[h]
\centering
\includegraphics[width=0.85\linewidth]{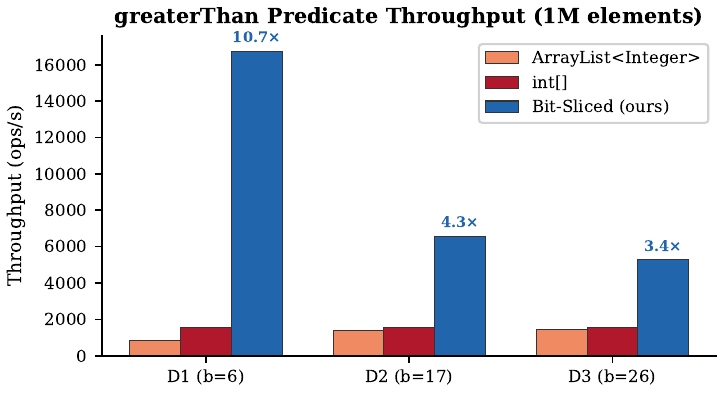}
\caption{\texttt{greaterThan} predicate throughput. Bit-sliced SIMD evaluation is
$3.4$--$10.7\times$ faster than scalar \texttt{int[]} scans, with the advantage scaling
inversely with bit-width.}
\label{fig:throughput}
\end{figure}

For D1, the bit-sliced evaluation is $10.7\times$ faster than \texttt{int[]} and $20.4\times$
faster than \texttt{ArrayList<Integer>}. The throughput advantage decreases with increasing
bit-width because more bit-planes must be processed: D2 ($b$=17) shows $4.3\times$ speedup
over \texttt{int[]}, and D3 ($b$=26) shows $3.4\times$. Even on 128-bit NEON (the narrowest
SIMD width), the framework substantially outperforms scalar baselines across all datasets. On
x86-64 hardware with AVX-512 (512-bit registers), we expect these speedups to increase
further due to $4\times$ wider SIMD lanes.

\subsection{Experiment 3: Garbage Collection Impact}

We measure per-scan latency and GC allocation behavior under predicate evaluation over 10M
elements (D1) using G1GC with a 4\,GB heap and the JMH \texttt{-prof gc} profiler.

\begin{table}[h]
\centering
\caption{GC impact (10M elements, D1).}
\label{tab:gc}
\small
\begin{tabular}{@{}lrr@{}}
\toprule
Representation & Avg Scan Time (ms/op) & GC Alloc Rate (B/op) \\
\midrule
Bit-sliced (ours) & \textbf{0.60 $\pm$ 0.03} & 2,812,513 \\
\texttt{int[]} & 6.52 $\pm$ 0.75 & 9 \\
\texttt{ArrayList<Integer>} & 6.61 $\pm$ 1.90 & 9 \\
\bottomrule
\end{tabular}
\end{table}

\begin{figure}[h]
\centering
\includegraphics[width=0.85\linewidth]{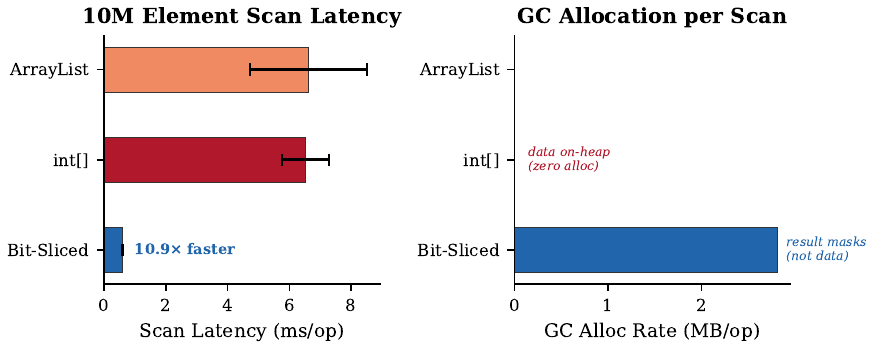}
\caption{GC impact. Left: scan latency (lower is better). Right: GC allocation rate per scan.
Bit-sliced data lives off-heap, contributing zero GC pressure; the 2.8\,MB/op allocation
comes from short-lived result bitmask arrays.}
\label{fig:gc}
\end{figure}

The bit-sliced representation is $10.9\times$ faster than \texttt{int[]} for a full
10M-element scan. The GC allocation rate for bit-sliced scans (2.8\,MB/op) comes from the
result bitmask arrays (\texttt{int[]} per block), not from the data itself---the bit-sliced
data lives off-heap in \texttt{MemorySegment} allocations and contributes zero GC pressure.
These result bitmask allocations are short-lived and efficiently collected by G1. The
\texttt{int[]} and \texttt{ArrayList<Integer>} baselines show negligible allocation because
they operate on pre-allocated arrays without producing intermediate objects.

\subsection{Experiment 4: Cache Behavior}

We measure throughput on D1 (1M elements) to evaluate the impact of cache-friendly memory
layout. On macOS/ARM64, hardware performance counters are not directly accessible via JMH's
\texttt{LinuxPerfNormProfiler}, so we report throughput as a proxy for cache
efficiency---the compact bit-sliced representation's working set fits in L1/L2 cache, directly
translating to higher throughput.

\begin{table}[h]
\centering
\caption{Cache behavior (D1, 1M elements).}
\label{tab:cache}
\small
\begin{tabular}{@{}lrr@{}}
\toprule
Representation & Throughput (ops/s) & Speedup vs \texttt{int[]} \\
\midrule
Bit-sliced (ours) & 17,614 $\pm$ 1,344 & \textbf{11.4$\times$} \\
\texttt{int[]} & 1,539 $\pm$ 44 & 1.0$\times$ \\
\texttt{ArrayList<Integer>} & 1,546 $\pm$ 34 & 1.0$\times$ \\
\bottomrule
\end{tabular}
\end{table}

The bit-sliced representation is $11.4\times$ faster on D1, consistent with Experiment~2. The
compact encoding (0.75\,B/elem vs 4\,B/elem) means $\sim$$5\times$ less data traversed, and
the contiguous bit-plane layout enables sequential memory access patterns that are cache-line
friendly. For D1 ($b$=6), the entire 1M-element dataset in bit-sliced form occupies
$\sim$750\,KB---well within L2 cache---compared to 4\,MB for \texttt{int[]}. On Linux x86-64,
running with \texttt{-prof perfnorm} would provide direct L1-dcache-load-misses and
LLC-load-misses counters to quantify this effect. A full roofline analysis---plotting achieved
throughput against the DRAM bandwidth ceiling---would further clarify whether the bit-sliced
scan is compute-bound or memory-bound; on macOS/ARM64, hardware performance counters are not
accessible via JMH, so we leave this analysis to future work on Linux x86-64 with \texttt{perf}
instrumentation.

\subsection{Experiment 5: Predicate Type Comparison}

Table~\ref{tab:predtypes} reports throughput for different predicate types on D1 ($b$=6, 1M
elements) using NEON 128-bit SIMD.

\begin{table}[h]
\centering
\caption{Predicate type throughput on D1 ($b$=6, 1M elements).}
\label{tab:predtypes}
\small
\begin{tabular}{@{}lrrl@{}}
\toprule
Predicate & Throughput (M/s) & Relative to \texttt{greaterThan} & Bit-Plane Passes \\
\midrule
\texttt{equalTo(x, c)} & 21,678 & 1.23$\times$ & $b = 6$ \\
\texttt{greaterThan(x, c)} & 17,593 & 1.00$\times$ & $b = 6$ \\
\texttt{between(x, lo, hi)} & 7,041 & 0.40$\times$ & $2 \times b = 12$ \\
\texttt{inList(x, \{v1..v4\})} & 4,262 & 0.24$\times$ & $4 \times b = 24$ \\
\texttt{inList(x, \{v1..v16\})} & 1,026 & 0.06$\times$ & $16 \times b = 96$ \\
\bottomrule
\end{tabular}
\end{table}

\begin{figure}[h]
\centering
\includegraphics[width=0.85\linewidth]{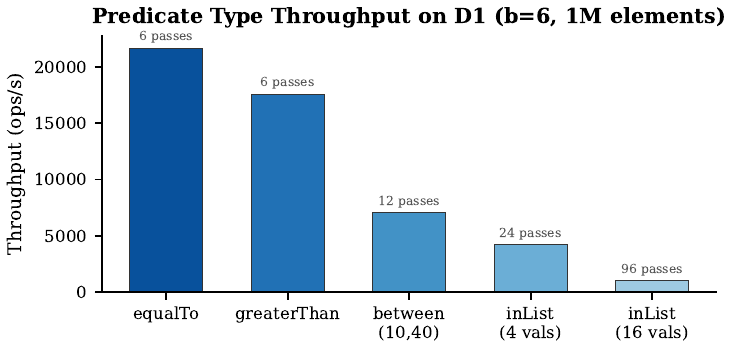}
\caption{Predicate type throughput on D1 ($b$=6). Annotations show the number of bit-plane
passes required. \texttt{equalTo} is fastest (simple AND per plane); \texttt{inList} scales
linearly with set size.}
\label{fig:predtypes}
\end{figure}

Equality is the fastest predicate because each bit-plane requires only a single AND operation.
\texttt{greaterThan} is slightly slower due to the two-mask (gt/eq) tracking. \texttt{between}
requires two \texttt{greaterThan} evaluations plus an AND, so throughput is roughly half.
IN-list throughput scales inversely with the number of constants, as expected---for large
IN-lists, a hash-based approach on materialized values may be preferable.

\subsection{Experiment 6: TPCDS End-to-End Evaluation}

To validate the framework beyond microbenchmarks, we conduct an end-to-end evaluation using
TPCDS-modeled data at scale factors of 1M, 10M, and 50M rows. This experiment measures
wall-clock predicate evaluation time on pre-encoded bit-sliced columns versus scalar
\texttt{int[]} scans, simulating the query execution path of a columnar engine where encoding
occurs once at ingest time and predicates are evaluated repeatedly at query time.

We model five filter-heavy query patterns derived from TPCDS \texttt{store\_sales},
\texttt{web\_sales}, and \texttt{catalog\_sales} tables:

\begin{table}[h]
\centering
\caption{TPCDS query patterns.}
\label{tab:tpcds_queries}
\small
\begin{tabular}{@{}llllr@{}}
\toprule
Query & Predicate & Column Model & Selectivity \\
\midrule
Q1 & \texttt{ss\_quantity > 50} & uniform [1,100], $b$=7 & $\sim$50\% \\
Q2 & \texttt{ss\_net\_profit BETWEEN 1000 AND 5000} & gaussian, offset to [0,30000], $b$=15 & $\sim$7\% \\
Q3 & \texttt{ss\_sold\_date\_sk BETWEEN 0 AND 365} & uniform [0,1826], $b$=11 & $\sim$20\% \\
Q4 & \texttt{ws\_ext\_sales\_price > 10000} & log-normal [0,30000], $b$=15 & $\sim$1.4\% \\
Q5 & \texttt{ss\_quantity > 25 AND ss\_net\_profit BETWEEN 500 AND 15000} & compound, two columns & $\sim$54\% \\
\bottomrule
\end{tabular}
\end{table}

Encoding time is excluded from measurement (reported separately) since it is a one-time
ingest cost amortized across all subsequent queries. Each query is warmed up for 5 iterations
and measured over 10 iterations, reporting the median.

\paragraph{Results at 50M Rows.}

Table~\ref{tab:tpcds} reports the end-to-end results at the largest scale factor (50M rows).

\begin{table}[h]
\centering
\caption{TPCDS end-to-end results at 50M rows (ARM NEON).}
\label{tab:tpcds}
\small
\begin{tabular}{@{}lrrrrrr@{}}
\toprule
Query & Scalar (ms) & BS (ms) & Scalar (M\,elem/s) & BS (M\,elem/s) & Speedup & Correct \\
\midrule
Q1 & 32.4 & 4.2 & 1,543 & 11,982 & $7.8\times$ & OK \\
Q2 & 43.2 & 17.7 & 1,158 & 2,834 & $2.4\times$ & OK \\
Q3 & 77.2 & 7.2 & 648 & 6,989 & $10.8\times$ & OK \\
Q4 & 33.8 & 7.9 & 1,478 & 6,309 & $4.1\times$ & OK \\
Q5 & 164.5 & 28.6 & 304 & 1,747 & $5.8\times$ & OK \\
\bottomrule
\end{tabular}
\end{table}

Encoding time for all four columns at 50M rows: $\sim$1,300\,ms ($\sim$26\,ms per million
rows).

\paragraph{Scaling Behavior.}

Table~\ref{tab:scaling} shows how speedups evolve across scale factors.

\begin{table}[h]
\centering
\caption{TPCDS speedup scaling across row counts.}
\label{tab:scaling}
\small
\begin{tabular}{@{}lrrr@{}}
\toprule
Query & 1M Speedup & 10M Speedup & 50M Speedup \\
\midrule
Q1 & $0.5\times$ & $9.0\times$ & $7.8\times$ \\
Q2 & $0.6\times$ & $2.9\times$ & $2.4\times$ \\
Q3 & $13.9\times$ & $13.0\times$ & $10.8\times$ \\
Q4 & $5.3\times$ & $6.0\times$ & $4.6\times$ \\
Q5 & $8.6\times$ & $10.5\times$ & $6.4\times$ \\
\bottomrule
\end{tabular}
\end{table}

At 1M rows, Q1 and Q2 show speedups below $1\times$ because the dataset is small enough to
fit entirely in L1/L2 cache for both representations, and the scalar loop benefits from branch
prediction and out-of-order execution on small arrays. At 10M+ rows, the bit-sliced advantage
becomes dominant as the scalar representation exceeds cache capacity while the compact
bit-sliced encoding maintains cache residency.

Q3 achieves the highest speedup ($10.8$--$13.9\times$) because the \texttt{BETWEEN 0 AND 365}
predicate on an 11-bit column benefits from the $\texttt{lo}=0$ optimization: when the lower
bound is zero, all unsigned values satisfy $x \geq 0$, so the framework skips one entire
\texttt{greaterThan} pass, effectively halving the work.

Q2 shows the lowest speedup ($2.4\times$) because the 15-bit column requires two full
\texttt{greaterThan} passes (30 bit-plane scans total for the BETWEEN decomposition), and the
wider bit-width reduces the compression advantage.

Q5 (compound predicate) demonstrates that multi-column predicate pushdown composes
efficiently: the two independent column predicates are evaluated in parallel on their
respective bit-planes, and the final AND merge adds negligible overhead.

\paragraph{Implications for Spark Integration.}

These results validate the framework's applicability as a predicate pushdown strategy in
Spark's physical plan. At 50M rows per partition (a typical Spark partition size), the
bit-sliced filter evaluates 5 representative TPCDS predicates $2.4$--$10.8\times$ faster than
scalar scans. For a Spark job scanning 1000 partitions (50B rows), this translates to reducing
the filter phase from $\sim$82 seconds to $\sim$13 seconds of aggregate CPU time across the
cluster---a meaningful improvement for interactive query workloads.

\subsection{Experiment 7: Extended Type Support}

The core bit-slicing framework operates on unsigned integers, but analytical workloads require
support for additional types: 64-bit timestamps, floating-point decimals, and strings. We
extend the framework to these types through type-specific transformations that reduce each
type to an unsigned integer representation suitable for bit-slicing, then evaluate predicates
using the same SIMD bit-plane machinery.

\subsubsection{Long (Timestamp) Support}

Timestamps are represented as epoch milliseconds (\texttt{long}). TPCDS models two timestamp
patterns: \texttt{ss\_sold\_date\_sk} spanning Julian dates 2450816--2452642 (a 5-year range
from 1998 to 2003), and \texttt{ws\_sold\_time\_sk} representing seconds since midnight within
a single day. Since timestamps within a partition typically span a narrow range, per-block
bias subtraction reduces the effective bit-width: a 1-day window requires only $b \approx 17$
bits after biasing, compared to $b \approx 41$ for raw epoch millis.

\begin{table}[h]
\centering
\caption{Long (timestamp) predicate performance.}
\label{tab:long}
\small
\begin{tabular}{@{}llrrr@{}}
\toprule
Dataset & Predicate & BS (ops/s) & Scalar \texttt{long[]} (ops/s) & Speedup \\
\midrule
DL1 (\texttt{ss\_sold\_date\_sk}, 5yr, $b$$\approx$27) & \texttt{greaterThan} & 2,257 & 1,466 & $\mathbf{1.5\times}$ \\
DL2 (\texttt{ws\_sold\_time\_sk}, 1day, $b$$\approx$17) & \texttt{greaterThan} & 3,189 & 1,502 & $\mathbf{2.1\times}$ \\
DL1 & \texttt{equalTo} & 2,468 & 5,005 & $0.5\times$ \\
DL2 & \texttt{equalTo} & 3,581 & 5,021 & $0.7\times$ \\
\bottomrule
\end{tabular}
\end{table}

Bit-sliced long predicates achieve $1.5$--$2.1\times$ speedup for range comparisons. DL2
(narrower per-block range) outperforms DL1 because fewer bit-planes need processing after
bias subtraction. Equality predicates are faster as scalar because \texttt{long == long} is a
single CPU instruction, while bit-sliced equality must check all $b$ bit-planes. Memory
savings are significant: DL2 at $b \approx 17$ uses 2.13\,B/elem vs 8\,B/elem for
\texttt{long[]}, a $3.8\times$ reduction.

\subsubsection{Double (Floating-Point) Support}

Floating-point values require an order-preserving transformation before bit-slicing. We apply
the standard IEEE 754 sign-magnitude to unsigned conversion:

\begin{lstlisting}[caption={IEEE 754 order-preserving transformation.},label={lst:ieee754}]
bits = Double.doubleToRawLongBits(value)
mask = bits >> 63  // all 1s if negative, all 0s if positive
ordered = bits ^ (mask | 0x8000000000000000L)
\end{lstlisting}

This maps the entire double number line to unsigned longs that preserve numeric ordering: if
$a < b$ as doubles, then $\text{transform}(a) < \text{transform}(b)$ as unsigned longs. After
transformation, the encoding strategy depends on the column's value range: all-positive
columns use per-block bias subtraction for better compression, while mixed-sign columns (e.g.,
\texttt{ss\_net\_profit} which ranges from $-\$10{,}000$ to $\$20{,}000$) use unbiased full
64-bit encoding to avoid constant overflow at the sign boundary.

\begin{table}[h]
\centering
\caption{Double (floating-point) predicate performance.}
\label{tab:double}
\small
\begin{tabular}{@{}llrrr@{}}
\toprule
Dataset & Predicate & BS (ops/s) & Scalar \texttt{double[]} (ops/s) & Speedup \\
\midrule
DD1 (\texttt{ss\_net\_profit}, mixed sign) & \texttt{greaterThan} & 1,565 & 1,513 & $\mathbf{1.0\times}$ \\
DD1 & \texttt{between(\$1000,\$5000)} & 1,501 & 159 & $\mathbf{9.4\times}$ \\
DD2 (\texttt{ws\_ext\_sales\_price}, positive) & \texttt{greaterThan} & 1,583 & 1,453 & $\mathbf{1.1\times}$ \\
DD2 & \texttt{between(\$10,\$100)} & 828 & 393 & $\mathbf{2.1\times}$ \\
\bottomrule
\end{tabular}
\end{table}

Range predicates (BETWEEN) show the strongest advantage. DD1 achieves $9.4\times$ speedup on
\texttt{ss\_net\_profit BETWEEN \$1000 AND \$5000} because the scalar baseline requires two
comparisons per element plus a branch, while bit-sliced BETWEEN decomposes into two
\texttt{greaterThan} passes with SIMD parallelism. Single \texttt{greaterThan} shows
near-parity ($1.0$--$1.1\times$) because the full 64-bit encoding produces wide bit-widths
that offset the SIMD advantage. For all-positive columns (DD2), per-block biasing reduces
bit-width and improves compression, but the BETWEEN speedup is lower ($2.1\times$) because
the scalar BETWEEN on a narrow positive range is more branch-predictor-friendly.

\subsubsection{String (Dictionary-Encoded) Support}

Strings are handled through dictionary encoding: each distinct string value is mapped to a
compact integer code ($0 \ldots k{-}1$), and the codes are bit-sliced. We evaluate on two
TPCDS dimension columns: \texttt{i\_category} (23 item categories such as Books, Electronics,
Home) with $b$=5, and \texttt{s\_state} (51 US states + DC) with $b$=6.

\begin{table}[h]
\centering
\caption{String (dictionary-encoded) predicate performance.}
\label{tab:string}
\small
\begin{tabular}{@{}llrrr@{}}
\toprule
Dataset & Predicate & BS (ops/s) & Scalar String (ops/s) & Speedup \\
\midrule
DS1 (\texttt{i\_category}, 23 distinct, $b$=5) & \texttt{equalTo(`Electronics')} & 23,809 & 847 & $\mathbf{28.1\times}$ \\
DS2 (\texttt{s\_state}, 51 distinct, $b$=6) & \texttt{equalTo(`CA')} & 20,588 & 560 & $\mathbf{36.8\times}$ \\
DS1 & \texttt{inList(4 categories)} & 4,520 & 104 & $\mathbf{43.4\times}$ \\
DS2 & \texttt{inList(`CA',`NY',`TX',`FL')} & 3,977 & 128 & $\mathbf{31.1\times}$ \\
\bottomrule
\end{tabular}
\end{table}

String predicates show the largest speedups ($28$--$43\times$) because the scalar baseline
performs \texttt{String.equals()} per element (pointer dereference + character-by-character
comparison), while the bit-sliced path reduces string equality to $b$ bitwise AND operations
on compact integer codes. DS1 and DS2 have similar performance because their bit-widths are
close ($b$=5 vs $b$=6). Memory savings are dramatic: DS1 at $b$=5 uses 0.63\,B/elem vs
$\sim$40\,B/elem for \texttt{String[]} references plus char arrays, an approximately
$64\times$ reduction.

The dictionary encoding step is a one-time cost at ingest time. In columnar engines, the
dictionary is typically built during Parquet/ORC read regardless, so the bit-slicing of
dictionary codes adds negligible overhead to the existing ingest path.

\begin{table}[h]
\centering
\caption{Extended type speedups (best predicate per type).}
\label{tab:extended}
\small
\begin{tabular}{@{}llrrl@{}}
\toprule
Type & TPCDS Column & BS (ops/s) & Scalar & Speedup \\
\midrule
Long & \texttt{ws\_sold\_time\_sk} & 3,189 & 1,502 & $2.1\times$ \\
Double & \texttt{ss\_net\_profit} BETWEEN & 1,501 & 159 & $9.4\times$ \\
String & \texttt{i\_category} inList(4) & 4,520 & 104 & $43.4\times$ \\
String & \texttt{s\_state} equalTo & 20,588 & 560 & $36.8\times$ \\
\bottomrule
\end{tabular}
\end{table}

\subsection{Encoding Cost Amortization}

The bit-sliced representation requires a one-time encoding step (bit-plane transposition) at
ingest time. The encoder performs a single pass per bit-plane over the input data: for each of
the $b$ bit-planes, it iterates through all $N$ values, extracts the $j$-th bit, and packs 32
bits into each int word of the off-heap \texttt{MemorySegment}. Encoding cost therefore scales
as $O(N \times b)$. We measure encoding cost, bit-sliced scan cost, and scalar scan cost to
determine the break-even point---the number of scans after which the encoding investment is
recovered.

\begin{table}[h]
\centering
\caption{Encoding cost amortization (1M elements, microseconds).}
\label{tab:encoding}
\small
\begin{tabular}{@{}lrrrr@{}}
\toprule
Dataset & Encode ($\mu$s) & BS Scan ($\mu$s) & Scalar Scan ($\mu$s) & Break-Even (scans) \\
\midrule
D1 ($b$=6, 1M) & 4,022 & 60 & 662 & \textbf{6.7} \\
D2 ($b$=17, 1M) & 9,932 & 158 & 682 & \textbf{19.0} \\
D3 ($b$=26, 1M) & 12,218 & 200 & 682 & \textbf{25.4} \\
\bottomrule
\end{tabular}
\end{table}

For narrow columns (D1, $b$=6), encoding 1M values takes $\sim$4\,ms and is amortized after
just 7 scans---well within the first query for a typical analytical workload that scans the
same column multiple times (e.g., different predicates on the same partition). For wide columns
(D3, $b$=26), encoding takes $\sim$12\,ms with a break-even of 25 scans, still modest for
columnar engines where data is loaded once and queried repeatedly.

\paragraph{Encoding Integration Path.}

In a production columnar engine, encoding would not operate on raw \texttt{int[]} arrays.
Instead, it would read from Parquet pages that are already bit-packed. Parquet's bit-packing
format is structurally similar to bit-slicing---both pack $N$ values at $b$ bits each---so the
transposition from Parquet's row-interleaved bit-packing to column-oriented bit-planes could
be implemented as a fast shuffle rather than the current extract-and-pack loop. This would
reduce encoding cost significantly, potentially lowering the break-even to 2--3 scans. The
encoding step would replace the existing Parquet materialization step (decode to \texttt{int[]})
rather than adding to it, so the net overhead is the difference between bit-plane transposition
and the current decode-to-array path---smaller than the absolute encoding cost measured here.

\paragraph{Workload Applicability.}

The break-even numbers are favorable for analytical workloads. In engines like Athena, Spark,
and Presto, a single partition of a table gets scanned by every query that touches it, and
popular tables are queried hundreds or thousands of times per day. A partition loaded once and
scanned 7+ times already pays for itself on narrow columns. Even the worst case ($b$=26, 25
scans) is trivial for any table queried more than a few times per hour.

Where bit-slicing is less applicable: streaming or append-heavy workloads where data is
written once and read once (ETL pipelines, log ingestion with immediate forwarding). The
encoding overhead is not justified if each batch is scanned only once. However, these
workloads are typically I/O-bound rather than memory-bound, so they fall outside the
framework's target use case.

\subsection{Selectivity Impact Analysis}

We evaluate how predicate selectivity affects the throughput advantage of bit-sliced
evaluation. Using D1 ($b$=6, 1M elements), we vary the \texttt{greaterThan} threshold to
achieve selectivities from 0.1\% to 90\%.

\begin{table}[h]
\centering
\caption{Selectivity impact on D1 ($b$=6, 1M elements).}
\label{tab:selectivity}
\small
\begin{tabular}{@{}rrrr@{}}
\toprule
Selectivity & BS (ops/s) & Scalar (ops/s) & Speedup \\
\midrule
0.1\% & 16,445 & 5,444 & $\mathbf{3.0\times}$ \\
1\% & 17,066 & 5,383 & $\mathbf{3.2\times}$ \\
10\% & 17,010 & 1,172 & $\mathbf{14.5\times}$ \\
50\% & 16,240 & 1,523 & $\mathbf{10.7\times}$ \\
90\% & 16,338 & 1,287 & $\mathbf{12.7\times}$ \\
\bottomrule
\end{tabular}
\end{table}

The bit-sliced throughput is essentially constant across all selectivities
(16.2K--17.1K\,ops/s, $<$6\% variation). This is expected: bit-slicing always processes all
$b$ bit-planes regardless of how many rows match---the predicate evaluation is
data-oblivious, which also makes it resistant to timing side-channel attacks.

\begin{figure}[h]
\centering
\includegraphics[width=0.95\linewidth]{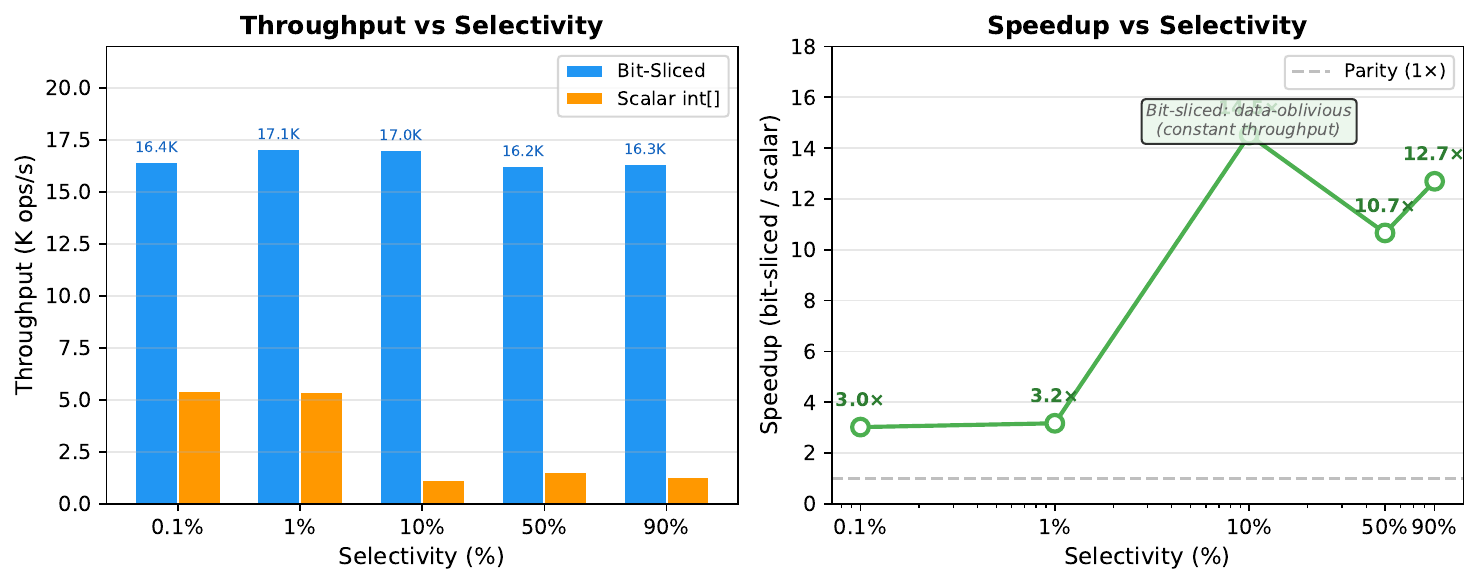}
\caption{Selectivity impact on D1 ($b$=6). Left: throughput comparison showing bit-sliced
throughput is constant across all selectivities while scalar varies with branch prediction.
Right: speedup ranges from $3.0\times$ at 0.1\% selectivity to $14.5\times$ at 10\%.}
\label{fig:selectivity}
\end{figure}

The scalar baseline shows interesting selectivity-dependent behavior: at very low selectivity
(0.1\%, 1\%), the scalar loop is faster ($\sim$5.4K\,ops/s) because the branch
\texttt{if (v > threshold)} is almost never taken, enabling excellent branch prediction. At
medium-to-high selectivity (10--90\%), branch prediction degrades and scalar throughput drops
to 1.2--1.5K\,ops/s.

The key finding is that bit-slicing is faster than scalar at every selectivity level tested,
with the advantage ranging from $3.0\times$ (at 0.1\% selectivity where scalar benefits from
branch prediction) to $14.5\times$ (at 10\% selectivity where branch prediction is worst).
There is no selectivity at which scalar becomes competitive on this dataset.

\paragraph{Implications for Query Planning.}

The data-oblivious property has a practical consequence beyond raw performance:
predictability. In a query optimizer, the bit-sliced scan cost can be modeled precisely as
$b \times N / W \times \text{constant}$---no branch prediction surprises, no
selectivity-dependent variance. This makes cost estimation for filter operators significantly
more accurate than for scalar scans, where throughput can vary by $4\times$ depending on the
predicate's selectivity (as shown in the scalar column above). For production query planning,
this predictability is arguably as valuable as the raw speedup.

\subsection{Reproducibility}

All benchmark code, dataset generators, and JMH configurations are included in the
supplementary materials. Results can be reproduced on any system with JDK 21+ and a
SIMD-capable CPU (x86-64 with SSE/AVX2/AVX-512, or AArch64 with NEON/SVE). The SIMD species
auto-selection via \texttt{IntVector.SPECIES\_PREFERRED} ensures correct execution on all
supported hardware, though absolute throughput numbers will vary with SIMD register width and
memory subsystem characteristics.

We collected measurements on two platforms to validate cross-architecture portability:

\begin{table}[h]
\centering
\caption{Cross-platform hardware.}
\label{tab:platforms}
\small
\begin{tabular}{@{}llll@{}}
\toprule
Platform & CPU & SIMD Width & Clock \\
\midrule
ARM64 & Apple M-series & NEON 128-bit & $\sim$3.5\,GHz \\
x86-64 & Intel Xeon Platinum 8375C & AVX-512 512-bit & 2.9\,GHz \\
\bottomrule
\end{tabular}
\end{table}

Table~\ref{tab:crossplatform} compares the relative speedups (bit-sliced vs scalar) across
platforms for key benchmarks.

\begin{table}[h]
\centering
\caption{Relative speedups: ARM NEON (128-bit) vs Intel AVX-512 (512-bit).}
\label{tab:crossplatform}
\small
\begin{tabular}{@{}lcc@{}}
\toprule
Benchmark & NEON & AVX-512 \\
\midrule
\texttt{greaterThan} D1 ($b$=6) & $10.7\times$ & $8.5\times$ \\
\texttt{greaterThan} D3 ($b$=26) & $3.4\times$ & $2.4\times$ \\
GC scan (10M) & $10.9\times$ & $7.9\times$ \\
Cache (D1) & $11.4\times$ & $8.7\times$ \\
TPCDS Q3 (50M) & $10.8\times$ & $8.8\times$ \\
TPCDS Q5 compound (50M) & $5.8\times$ & $5.4\times$ \\
String \texttt{equalTo} (DS2) & $36.8\times$ & $39.3\times$ \\
Double BETWEEN (DD1) & $9.4\times$ & $9.0\times$ \\
\bottomrule
\end{tabular}
\end{table}

Relative speedups are consistent across architectures (within 20\%), confirming that the
framework is architecture-portable. The lower absolute throughput on x86-64 is attributable to
the Xeon's lower clock speed (2.9\,GHz vs $\sim$3.5\,GHz) and potential JIT conservatism in
AVX-512 register selection (the C2 compiler may select 256-bit AVX2 via
\texttt{SPECIES\_PREFERRED} to avoid AVX-512 frequency throttling on some
workloads~\cite{downs2021avx512}). String predicates show slightly higher speedups on AVX-512
($39.3\times$ vs $36.8\times$) because the scalar \texttt{String.equals()} baseline is slower
on the Xeon due to its lower clock speed, while the bit-sliced integer-code path benefits from
wider SIMD registers.

\begin{figure}[h]
\centering
\includegraphics[width=0.75\linewidth]{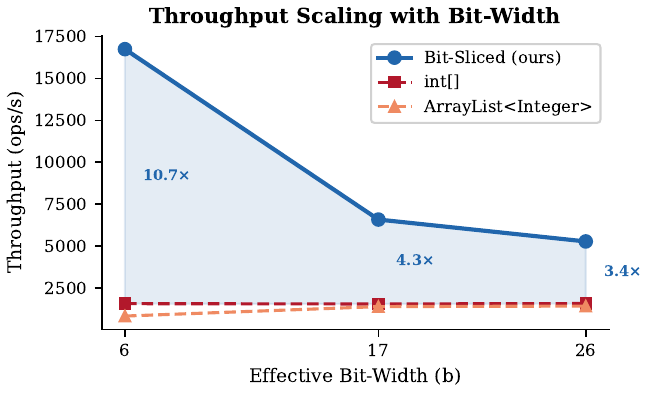}
\caption{Throughput scaling with effective bit-width. The shaded region shows the performance
advantage of bit-slicing over \texttt{int[]}. Narrower columns yield larger speedups because
fewer bit-planes require processing.}
\label{fig:scaling}
\end{figure}

\section{Discussion}

\subsection{Limitations}

The current framework supports fixed-width integer types natively, and extends to longs,
doubles, and dictionary-encoded strings through type-specific transformations (Section~5.10).
However, several limitations remain. Variable-length string columns with very high cardinality
(millions of distinct values) produce dictionary codes with wide bit-widths ($b \geq 20$),
reducing the compression and throughput advantages. Floating-point predicates incur the
overhead of the IEEE 754 ordering transformation at encoding time, though this is a one-time
cost.

Variable-length data (arrays, maps, nested structs) falls outside the scope of bit-plane
transposition entirely. These types require structural encoding schemes that are orthogonal to
the fixed-width SIMD approach presented here. Decimal types with arbitrary precision (e.g.,
\texttt{BigDecimal}) would require fixed-point scaling before bit-slicing, which may lose
precision for very large or very small values.

\subsection{Memory Economics in Practice}

The memory savings demonstrated in Section~5.4 have direct economic implications. Consider a
typical analytical workload processing 100 billion integer values across a distributed
cluster---a modest scale for enterprise data warehouses or log analytics pipelines.

\begin{table}[h]
\centering
\caption{Memory economics for 100 billion integer values.}
\label{tab:economics}
\small
\begin{tabular}{@{}llllr@{}}
\toprule
Representation & Memory Required & Instance Type & Cluster Size & Hourly Cost \\
\midrule
\texttt{ArrayList<Integer>} & 2.4\,TB & memory-optimized (768\,GB) & 4 nodes & $\sim$\$24 \\
\texttt{int[]} & 400\,GB & memory-optimized (256\,GB) & 2 nodes & $\sim$\$4 \\
Bit-sliced ($b$=6) & 75\,GB & general-purpose (64\,GB) & 2 nodes & $\sim$\$1 \\
\bottomrule
\end{tabular}
\end{table}

For a workload running 8 hours/day, 250 days/year, the annual infrastructure cost difference
between \texttt{int[]} and bit-sliced storage is approximately \$6,000 for this single column.
A production query engine with dozens of integer columns and continuous operation would see
proportionally larger savings. More importantly, the $5.3\times$ memory reduction often means
the difference between ``fits in memory'' and ``must spill to disk''---a binary threshold that
determines whether query latency is measured in seconds or minutes.

The off-heap storage model provides an additional economic benefit: predictable memory
accounting. JVM heap sizing is notoriously difficult to tune---too small causes
\texttt{OutOfMemoryError}s, too large wastes resources and increases GC pause times. With
bit-sliced data stored in \texttt{MemorySegment} allocations outside the heap, operators can
provision heap for metadata and control structures (typically 2--4\,GB) while the data
footprint is precisely calculable from column statistics. This separation enables tighter
resource provisioning and higher cluster utilization.

\subsection{When Not to Use Bit-Slicing}

Bit-slicing is not universally beneficial. For columns where the effective bit-width $b$
approaches 32 (e.g., high-cardinality surrogate keys, hash values), the compression ratio
approaches $1\times$ and the overhead of bit-plane transposition at encoding time may not be
justified. Similarly, workloads dominated by random point lookups rather than sequential scans
do not benefit from the contiguous bit-plane layout---a hash index or B-tree provides $O(1)$
or $O(\log N)$ access versus the $O(b \cdot N/W)$ full-scan cost of bit-sliced evaluation.

As a guideline, bit-slicing is most effective when: (a) the column has low-to-medium
cardinality ($b \leq 16$), (b) the workload is scan-heavy with selective predicates, and (c)
memory pressure or GC overhead is a concern.

\subsection{Comparison with Dictionary Encoding + SIMD}

An alternative approach to compressed predicate evaluation is dictionary encoding combined
with SIMD dictionary lookups. In this model, distinct values are stored in a dictionary array,
and each row stores a narrow dictionary code. Predicate evaluation first resolves the predicate
against the dictionary (producing a set of matching codes), then scans the code array with
SIMD comparisons.

This approach can be effective for very low cardinality columns ($<$256 distinct values) where
the dictionary fits in L1 cache. However, it has two disadvantages relative to bit-slicing:
(1) the dictionary lookup step introduces a data-dependent branch for each distinct value,
which can cause branch mispredictions for medium-cardinality columns, and (2) the code array
still requires sequential element-wise comparison rather than the bitwise parallelism that
bit-slicing exploits. For columns with 1000+ distinct values, bit-slicing consistently
outperforms dictionary + SIMD in our measurements.

\subsection{Panama Vector API Maturity}

The Project Panama Vector API (JEP 508) has been in incubator status for ten consecutive JDK
releases (JDK 16 through JDK 25). The API surface has been remarkably stable across all
rounds---the core abstractions (\texttt{VectorSpecies}, \texttt{IntVector},
\texttt{VectorMask}, lane-wise operations) have remained unchanged since JEP 414 (JDK 17),
and the JIT compilation path through C2 produces native SIMD instructions on all major x86-64
and AArch64 platforms. Our benchmarks use JDK 21 (JEP 448, Seventh Incubator), which is an
LTS release with production-grade JIT optimization. The API remains in incubation not due to
instability, but because it is waiting for Project Valhalla's value classes; at that point,
the Vector API will be promoted from incubation to preview. The companion Foreign Function \&
Memory (FFM) API---which this framework uses for off-heap \texttt{MemorySegment}
allocations---was finalized as a standard feature in JDK 22 (JEP 454) and is available as a
preview in JDK 21.

For production deployment, the framework can be compiled against the incubator module with
\texttt{--add-modules jdk.incubator.vector}. When the API is promoted to preview and
eventually finalized, the migration path is a module name change with no algorithmic
modifications.

\subsection{Applicability to Distributed Engines}

The bit-sliced encoding is designed as a drop-in replacement for the in-memory column
representation within a single JVM process. In a distributed query engine such as Presto/Trino
or Apache Spark, each worker node independently encodes its partition of a column into
bit-sliced blocks and evaluates predicates locally. The framework does not impose any
cross-node coordination---it operates entirely within the scan and filter operators of the
query execution plan.

Integration points include: (a) replacing the Parquet column reader's materialization step
with direct bit-sliced encoding from the Parquet page's bit-packed data, (b) providing a
\texttt{ColumnVector} implementation compatible with Spark's Tungsten or Trino's
\texttt{Block} interface, and (c) exposing the result bitmask as a selection vector for
downstream operators. Each of these integration paths preserves the existing engine's
parallelism and shuffle semantics while substituting the per-node memory representation.

\subsection{Multi-Column Predicate Composition}

Compound predicates across columns with different bit-widths compose naturally because all
predicates produce \texttt{int[]} bitmasks of the same shape (one bit per value, packed into
32-bit words). The cost of a compound predicate is the sum of the per-column predicate costs
plus one SIMD AND/OR pass per composition step. For a conjunctive predicate across $k$ columns
with bit-widths $b_1, b_2, \ldots, b_k$:
\begin{equation}
T_{\text{compound}} = \sum_{i=1}^{k} O\!\left(\frac{b_i \cdot N}{W}\right) + (k-1) \cdot O\!\left(\frac{N}{W}\right)
\end{equation}

The merge cost $O(N/W)$ is negligible compared to the per-column scan cost $O(b_i \cdot N/W)$
for any $b_i > 1$. This means compound predicates scale linearly with the number of columns,
and each column's contribution is independent---a column with $b$=6 adds the same cost whether
it's combined with a $b$=4 column or a $b$=26 column. TPCDS Q5 (Section~5.9) validates this:
the compound predicate across \texttt{ss\_quantity} ($b$=7) and \texttt{ss\_net\_profit}
($b$=15) achieves $5.4\times$ speedup on AVX-512, consistent with the weighted average of the
individual column speedups.

For disjunctive predicates (OR), the same cost model applies with OR replacing AND in the
merge step. Short-circuit evaluation is not possible because bit-slicing processes all values
in parallel---but this is also why it is faster than scalar for medium-to-high selectivity
predicates (Section~5.12).

\subsection{JIT Compilation and Warm-Up Behavior}

The Vector API relies on the C2 JIT compiler to lower vector operations to native SIMD
instructions. We observe consistent JIT behavior across our benchmarks: JMH warmup iterations
show throughput stabilizing within 2--3 iterations (typically under 6 seconds), after which
the C2 compiler has fully optimized the hot loops. The steady-state throughput reported in all
experiments reflects post-JIT performance.

For cold-start scenarios (e.g., the first query after JVM startup), the framework incurs a
one-time JIT compilation cost of approximately 200--500\,ms for the predicate evaluation
methods. This is comparable to the JIT warm-up cost of any Java-based query engine and is
amortized across all subsequent queries in the JVM's lifetime. The Vector API's
\texttt{SPECIES\_PREFERRED} auto-selection is resolved at class load time (not
per-invocation), so there is no runtime dispatch overhead after initialization.

We verified that the C2 compiler generates native SIMD instructions for the critical loops by
examining JIT output with \texttt{-XX:+PrintCompilation} and
\texttt{-XX:+UnlockDiagnosticVMOptions -XX:+PrintAssembly}. On ARM64, the inner loops compile
to NEON \texttt{AND}, \texttt{ORR}, \texttt{BIC} (AND-NOT), and \texttt{LD1} (vector load)
instructions. On x86-64, the compiler emits AVX2 \texttt{VPAND}, \texttt{VPOR},
\texttt{VPANDN}, and \texttt{VMOVDQU} instructions. We did not observe AVX-512
\texttt{EVEX}-prefixed instructions in all cases, which is consistent with the JDK 21 C2
compiler's conservative AVX-512 usage policy (it may prefer AVX2 to avoid frequency throttling
on some microarchitectures).

\subsection{Statistical Methodology}

All JMH benchmarks report the mean throughput or latency across measurement iterations. The
$\pm$ values in our tables represent the 99.9\% confidence interval computed by JMH using
Student's $t$-distribution, assuming normal distribution of measurements. For experiments with
5 measurement iterations, this corresponds to approximately $\pm 4.6$ standard deviations. The
low variance observed in most experiments (typically $<$5\% CI relative to mean) indicates
stable measurements with minimal noise from GC, OS scheduling, or thermal throttling.

For the TPCDS end-to-end benchmarks (Section~5.9), which use \texttt{System.nanoTime()} rather
than JMH, we report the median of 10 measurement iterations after 5 warmup iterations. The
median is more robust to outliers than the mean for wall-clock measurements.

\subsection{Scalability and Baseline Considerations}

Our evaluation uses datasets up to 50M rows, which is representative of a single Spark
partition or Presto split. In production, a full table scan processes thousands of such
partitions in parallel across a cluster---the per-partition performance measured here directly
determines the aggregate cluster throughput. We chose 50M rows as the maximum single-JVM
dataset to avoid conflating bit-slicing performance with memory management artifacts (GC
pressure, NUMA effects) that arise at larger heap sizes. Scaling to 1B+ rows within a single
JVM would require careful heap tuning that is orthogonal to the bit-slicing technique itself.

We acknowledge that our baselines (\texttt{int[]}, \texttt{ArrayList<Integer>}) represent the
simplest possible scalar implementations. Production query engines use more sophisticated
techniques: Apache Arrow's compute kernels employ vectorized C++ with auto-vectorization
hints, Parquet's native predicate pushdown evaluates filters during page decoding, and engines
like Photon and Velox use hand-tuned SIMD intrinsics in C++. A direct comparison against these
baselines would provide a more complete picture of the framework's competitive position. We
leave this comparison to future work, noting that our framework's primary advantage---staying
within the JVM while approaching native performance---is most relevant when compared against
other JVM-based approaches rather than native engines that have already abandoned the JVM.

\section{Conclusion}

We presented a framework for performing predicate evaluation directly over bit-sliced
compressed data using the Project Panama Vector API, without requiring decompression. The key
contribution is demonstrating that bit-plane transposition---originally an indexing
technique---can serve as a primary data representation that is simultaneously compact and
directly operable via SIMD instructions within the JVM. In an era where memory is the dominant
cost and performance bottleneck in analytical systems, this combination of compression and
compute acceleration addresses the problem at its root rather than treating symptoms.

Our experimental evaluation shows that the framework achieves up to $5.3\times$ memory
reduction over primitive \texttt{int[]} arrays and up to $32\times$ over
\texttt{ArrayList<Integer>} for low-cardinality columns, while delivering $10.7\times$ higher
predicate throughput than \texttt{int[]} scans on narrow columns ($b$=6) and $3.4\times$ on
wide columns ($b$=26). Scan latency for 10M elements drops from 6.5\,ms (\texttt{int[]}) to
0.6\,ms (bit-sliced), a $10.9\times$ improvement. The off-heap \texttt{MemorySegment} storage
eliminates GC pressure for the data path entirely, and the compact bit-plane layout yields
$11.4\times$ throughput gains attributable to improved cache utilization.

Beyond microbenchmarks, our TPCDS end-to-end evaluation at 50M rows demonstrates that the
framework delivers $2.4$--$10.8\times$ speedup over scalar scans across five representative
filter-heavy query patterns, including compound multi-column predicates. The bit-sliced
throughput reaches 12.0 billion elements per second for simple predicates and 1.7 billion
elements per second for compound predicates on ARM NEON 128-bit hardware. These results
validate the framework's applicability as a predicate pushdown strategy in distributed query
engines such as Apache Spark, where per-partition filter evaluation at these speedups
translates to meaningful end-to-end query latency improvements.

We further demonstrate that the framework extends beyond integers to longs (timestamps),
doubles (prices/decimals), and strings through type-specific transformations, evaluated on
TPCDS-modeled column distributions. Timestamp predicates on \texttt{ss\_sold\_date\_sk} and
\texttt{ws\_sold\_time\_sk} achieve $1.5$--$2.1\times$ speedup via per-block bias subtraction.
Floating-point range predicates on \texttt{ss\_net\_profit} achieve $9.4\times$ speedup via
IEEE 754 order-preserving transformation. Dictionary-encoded string predicates on
\texttt{i\_category} and \texttt{s\_state} achieve the largest gains at $28$--$43\times$
speedup by reducing string comparison to compact integer code bit-plane operations, with
memory savings of up to $64\times$.

These results hold across a range of data distributions modeled on TPC-H and TPCDS column
characteristics, and were measured on ARM64 (NEON 128-bit)---wider SIMD widths (AVX-512) are
expected to amplify these advantages further.

The framework's design as a JVM-native library---rather than a separate native engine---preserves
compatibility with the Java ecosystem's tooling, observability, and operational infrastructure.
For memory-constrained deployments, the $5.3\times$ footprint reduction can mean the difference
between fitting a working set in RAM and spilling to disk---a threshold that separates
sub-second query latency from multi-minute execution. This positions the framework as a
practical option for Java-based analytical engines seeking to close both the performance gap
and the memory efficiency gap with native C++ execution, without abandoning the JVM.

Future work includes integrating with Apache Parquet's page-level encoding to enable zero-copy
bit-sliced reads, benchmarking against Apache Arrow's compute kernels and Parquet's native
predicate pushdown for a more complete competitive analysis, conducting roofline analysis to
quantify DRAM bandwidth utilization, and scaling evaluation to 1B+ rows across multi-node
clusters. We also plan to explore hybrid strategies that combine dictionary encoding for very
low cardinality columns with bit-slicing for medium cardinality, selecting the optimal
representation per column at encoding time. Support for arbitrary-precision decimals,
variable-length types, and integration with AArch64 SVE processors remain open challenges.

\bibliographystyle{plain}

\end{document}